\documentclass[pre,aps,showpacs,nofootinbib,twocolumn]{revtex4}
\usepackage{amssymb}

\usepackage{graphicx}

\begin{document}

\author{$^{1}$\textbf {M. L. Kuli\'{c}} and $^{2}$\textbf {O. V. Dolgov}}
\title{\textbf {Why is the ARPES anti-nodal singularity at $40$ $meV$
shifted in superconducting state of HTSC, but the kink at $70$
$meV$ is not?}}

\begin{abstract}
The theoretical model for the quasiparticle self-energy $\Sigma
{\bf (k},\omega )$ in HTSC is proposed, which is based on the
forward scattering peak in the electron-phonon (EPI) interaction.
By assuming that EPI dominates, the model explains qualitatively
and in a consistent way the recent ARPES results. The latter show
a kink in the normal state quasiparticle energy at 70 meV in the
nodal direction, which is (surprisingly) {\em not shifted%
} in the superconducting (SC) state, while the singularity at 40
meV in the anti-nodal direction is shifted by the SC gap. The
model predicts a dip-hump structure in the spectral function
$A({\bf k}_{F},\omega )$, which is observed in ARPES.
\end{abstract}

\bigskip
\address{$^{1}$Institute for Physics, Theory II, University Augsburg,
86135 Augsburg, Germany\\ $^{2}$Institute for Theoretical Physics,
University T\"{u}bingen, T\"{u}bingen, Germany}
\date{\today }

\maketitle

{\it Introduction} - The pairing mechanism in high-temperature
superconductors (HTSC) is still under the debate \cite{allen2}.
However, recent ARPES \cite {lanzara}, \cite{damascelli}, gives
evidence for pronounced phonon effects in the quasiparticle
energy, while the theory \cite{kulic2}, \cite{kulic} predicts that
strong correlations give rise to a pronounced forward scattering
peak (FSP) in the electron-phonon interaction (EPI) and in the
non-magnetic impurity scattering - {\em the FSP model}. It
predicts: (i) d-wave pairing is due to the EPI and the residual
Coulomb repulsion, which triggers it; (ii)
the transport coupling constant $\lambda _{tr}$ (entering the resistivity, $%
\varrho \sim \lambda _{tr}T$) is much smaller than the pairing one $\lambda $%
, i.e. $\lambda _{tr}\ll \lambda $; (iii) robustness of d-wave
pairing in the presence of non-magnetic impurities, etc. The FSP
in the EPI of strongly correlated systems is a general effect
which {\em affects electronic coupling to all phonons}. Numerical
calculations on the Hubbard model with the EPI \cite{hanke}
confirme the theory of Ref. \cite{kulic2}, \cite{kulic}.

Recent ARPES on various HTSC families \cite{lanzara} show a {\em kink%
} in the normal (N)\ state quasiparticle spectrum, $\omega
(\xi_{\bf k})$, in the nodal direction $(0,0)-(\pi ,\pi )$ at
energy $\omega _{kink}^{(70)}\lesssim 70$ $meV$, which is a
characteristic oxygen vibration energy, i.e. $\omega
_{kink}^{(70)}\sim \omega
_{ph}^{(70)}$. Surprisingly the kink is {\it %
not shifted} in the SC state, contrary to the standard Eliashberg
theory \cite{schrieffer}. Furthermore, ARPES on
$La_{2-x}Sr_{x}CuO_{4}$ and $BISCO$ crystals \cite {shen2} show
that in the anti-nodal direction $(\pi ,0)-(\pi ,\pi )$ a
singularity appears in $\omega
(\xi_{\bf k})$ in the N state ($T>T_{c}$) at $%
\omega _{sing}^{(40)}\approx 40$ $meV$, which is a also a
characteristic oxygen vibration energy $\omega _{sing}^{(40)}\sim
\omega _{ph}^{(40)}$. It is {\em shifted in the SC state} (at low
$T$) to $\omega \approx
60$ $meV(=\omega _{ph}^{(40)}+\Delta_{0})$, where $\Delta_{0}%
(\approx 20$ $meV)$ is approximately the maximal SC gap at the
anti-nodal point. The different shifts of $\omega _{kink}^{(70)}$
and $\omega _{sing}^{(40)}$ in the SC state we call the {\em
ARPES} {\em shift-puzzle}.

Why is the anti-nodal singularity $\omega _{sing}^{(40)}$ shifted
in the SC state, but the nodal kink $\omega _{kink}^{(70)}$ is
not? The {\em ARPES shift-puzzle} can not be explained by the
standard (with the integration also over the whole Fermi surface)
Eliashberg (or BCS)
theory for any kind of pairing \cite{schrieffer}, which predicts that $%
\omega _{sing}^{(40)}$ and $\omega _{kink}^{(70)}$ are shifted in
the SC state by the same value $\Delta _{0}$, i.e. $\omega
_{sing}^{(40)}\rightarrow \omega _{ph}^{(40)}+\Delta _{0}$ and
$\omega _{kink}^{(70)}\rightarrow \omega _{ph}^{(70)}+\Delta
_{0}$, where  $\Delta _{0}$ is the maximal gap value. ARPES can
not be explained by the spin-fluctuation theory (SF) based on the
$41$ $meV$ magnetic-resonance mode \cite{norman} because of at
least two reasons: (i) the kink at $70$ $meV$ is present also in
the N state, where there is no magnetic resonance mode and (ii)
the kink is seen in $La_{2-x}Sr_{x}CuO_{4}$, where there is no
magnetic resonance mode neither in the N nor in SC state
\cite{lanzara}. ARPES gives also evidence for the linear (in
$\omega$)
contribution to $\mathop{\rm Im}%
\Sigma (\omega )$ due to the Coulomb interaction (SF is only part
of it) \cite{damascelli}, i.e. $\mathop{\rm Im}%
\Sigma_{C} (\omega )\sim -\pi \lambda _{C}\omega /2$ for $T<\omega
<\Omega _{C}$. It is clearly discernable in ARPES for $\omega_{ph}
<\omega <\Omega _{C}$ with $\lambda _{C}\lesssim 0.4$. ARPES
\cite{lanzara}, \cite{damascelli} gives also that $\lambda
_{ph}>1$.

Here we show that the {\em ARPES shift-puzzle} implies {\em the
FSP model} with the following ingredients: {\it (i)}{\em \ }the
EPI is dominant \cite{kulic} and its spectral function $\alpha
^{2}F({\bf k},{\bf k}^{\prime },\Omega )$ has a pronounced FSP (at
${\bf k}-{\bf k}^{\prime }=0$) due to strong correlations. Its
width is very narrow $\mid {\bf k}-{\bf k}^{\prime }\mid _{c}\ll
k_{F}$ even for overdoped systems; {\it (ii)} the dynamical part
(beyond the Hartree-Fock) of
the Coulomb interaction is characterized by the spectral function $S_{C}(%
{\bf k},{\bf k}^{\prime },\Omega )$. The {\em ARPES shift-puzzle}
implies that $S_{C}$ is {\it either peaked} at small $\mid {\bf
k}-{\bf k}^{\prime }\mid$, {\it or it is so small} that it does
not affect the shift. Which of these possibilities is realized is
a matter of future ARPES. In order to minimize numerical
calculations we assume here that the former case is realized.;
{\it (iii)} The scattering potential on non-magnetic impurities
has pronounced FSP, due to strong correlations \cite{kulic2},
\cite{kulic}.
In the following we calculate $\Sigma (%
{\bf k},\omega )$ in the $N$ and $SC$ state and show that the
anti-nodal singularity at $\omega _{sing}^{(40)}$ is shifted in
the $SC$ state by $\Delta _{0}$, while the nodal kink at $\omega
_{kink}^{(70)}$ is not. The FSP model predicts also the existence
of a dip-hump structure in $A({\bf k},\omega )$.

{\it Eliashberg equations for the FSP model - }The normal and the anomalous
Matsubara Green's functions are defined \cite{allen} by $G(k)=-[Z(k)i\omega_{n} +%
\bar{\xi}(k)]/D(k)$ and $F(k)=Z(k)\Delta (k)/D(k)$ ($k=({\bf
k},\omega
_{n})$), respectively, where $D(k)=(Z(k)\omega _{n})^{2}+[\bar{\xi}%
^{2}(k)+(Z(k)\Delta (k))^{2}]$. The diagonal odd part of the self energy is $%
\Sigma ({\bf k},\omega _{n})=i\omega _{n}[1-Z({\bf k},\omega_{n} )](=-\Sigma (%
{\bf k},-\omega _{n}))$, while its even part is $\Sigma _{e}({\bf
k},\omega _{n})=\bar{\xi}(k)-\xi _{0}({\bf k})(=\Sigma _{e}({\bf
k},-\omega _{n}))$, where $\xi _{0}({\bf k})=\epsilon _{0}({\bf
k})-\mu $. Since in the following we assume the electron-hole
symmetry, then $\Sigma _{e}({\bf k},\omega _{n})\approx
\bar{\xi}({\bf k})-\xi _{0}({\bf k})$, i.e. it is a dull function
of $\omega $ which renormalizes the chemical potential and the
bare quasiparticle energy \cite {kulic}, \cite{allen}. The 2D
Fermi surface
of HTSC oxides is parametrized by ${\bf k}=(k_{F}+k_{\perp },k_{F}\varphi )$%
, where $k_{F}(\varphi )$ is the Fermi momentum and $k_{F}\varphi
$ is the tangential component of ${\bf k}$ at the point on the
Fermi surface \cite {allen}. In that case $\xi ({\bf k})\approx
v_{F}(\varphi )k_{\perp }$ and $\int d^{2}k[...]\approx \int \int
d\xi k_{F}(\varphi )d\varphi /v_{F}(\varphi )=\int \int N_{\varphi
}(\xi )d\xi d\varphi $. For simplicity we assume that near the
Fermi surface the EPI spectral function $\alpha
_{ph}^{2}F({\bf k,k}^{\prime },\Omega )$ is weakly dependent on energies $%
\xi ,\xi ^{\prime }$, i.e. $\alpha _{ph}^{2}F({\bf k,k}^{\prime
},\Omega )\approx \alpha _{ph}^{2}F(\varphi {\bf ,}\varphi
^{\prime },\Omega )$ \cite {allen} - see the item ({\bf i}) in the
discussion. In the presence of strong correlations one has $\alpha
_{ph}^{2}F(\varphi {\bf ,}\varphi ^{\prime },\Omega )\sim \gamma
_{c}^{2}(\varphi {\bf -}\varphi ^{\prime }),$ where the charge vertex $%
\gamma _{c}(\varphi {\bf -}\varphi ^{\prime })$ is strongly peaked at $%
\delta \varphi (=\varphi {\bf -}\varphi ^{\prime })=0$ with the width $%
\delta \varphi _{w}\ll \pi $ even for overdoped hole doping
\cite{kulic2}, \cite{kulic}. Then in the leading order one has
$\alpha _{ph}^{2}F(\varphi {\bf ,}\varphi ^{\prime },\Omega
)\approx \alpha _{ph}^{2}F(\varphi ,\Omega )\delta (\varphi {\bf
-}\varphi ^{\prime })$ which picks up the main physics
\cite{kulic} whenever $\delta \varphi _{w}\ll \pi $ - see also the
item ({\bf i}) in the discussion. After integration over $\xi
^{\prime }$ and for $N_{\varphi }(\xi )\approx N_{\varphi }(0)$
one obtains the Eliashberg equations

\[
\tilde{\omega}_{n,\varphi }=\omega _{n}+\pi T\sum_{m}\frac{\lambda
_{ph,\varphi }(\omega _{n}-\omega _{m})\tilde{\omega}_{m,\varphi }}{\sqrt{%
\tilde{\omega}_{m,\varphi }^{2}+\tilde{\Delta}_{m,\varphi }^{2}}}
\]
\begin{equation}
+\Sigma _{C,n,\varphi }+\frac{\gamma _{1,\varphi }\tilde{\omega}_{n,\varphi }%
}{\sqrt{\tilde{\omega}_{n,\varphi }^{2}+\tilde{\Delta}_{n,\varphi }^{2}}}
\eqnum{1}
\end{equation}

\[
\tilde{\Delta}_{n,\varphi }=\pi T\sum_{m}\frac{\lambda _{ph,\varphi }(\omega
_{n}-\omega _{m})\tilde{\Delta}_{m,\varphi }}{\sqrt{\tilde{\omega}%
_{m,\varphi }^{2}+\tilde{\Delta}_{m,\varphi }^{2}}}+\tilde{\Delta}%
_{C,n,\varphi },
\]
\begin{equation}
+\frac{\gamma _{2,\varphi }\tilde{\Delta}_{n,\varphi }}{\sqrt{\tilde{\omega}%
_{n,\varphi }^{2}+\tilde{\Delta}_{n,\varphi }^{2}}}-\pi
T\sum_{m}^{\omega _{c}}\int d\varphi ^{\prime }\frac{N_{\varphi
^{\prime }}(0)}{N(0)}\frac{\mu _{\varphi {\bf ,}\varphi ^{\prime
}}^{\ast }\tilde{\Delta}_{m,\varphi
^{\prime }}}{\sqrt{\tilde{\omega}_{m,\varphi ^{\prime }}^{2}+\tilde{\Delta}%
_{m,\varphi ^{\prime }}^{2}}},  \eqnum{2}
\end{equation}
where the EPI\ coupling $\lambda _{ph,\varphi }(\omega _{n}-\omega
_{m})$ is given by
\begin{equation}
\lambda _{ph,\varphi }(\omega _{n}-\omega _{m})=2\int_{0}^{\infty }d\Omega
\frac{\alpha _{ph,\varphi }^{2}F_{\varphi }(\Omega )\Omega }{\Omega
^{2}+(\omega _{n}-\omega _{m})^{2}}.  \eqnum{3}
\end{equation}
$\Sigma _{C,n,\varphi }$ is due to the dynamical Coulomb effects
and it is the most difficult part of the problem. The theory
predicts $\Sigma _{C}\sim (\Gamma
_{e}/\varepsilon _{e})V_{C}G$ where $V_{C}$ is the Coulomb potential, $%
\varepsilon _{e}\neq 1$ is the electronic dielectric function and
$\Gamma _{e}$ is the vertex function \cite{allen}. The {\em ARPES
shift-puzzle} implies that $(\Gamma _{e}/\varepsilon _{e})V_{C}$
should be either peaked at small angles or so small that it does
not affect the energy-shift. The former is also supported by the
theory of strongly correlations where $\Gamma _{e}=\gamma
_{c}(\varphi {\bf -}\varphi ^{\prime })[1+...]$ and $\gamma
_{c}(\varphi {\bf -}\varphi ^{\prime })$ is peaked at small angles
\cite{kulic2}, \cite{kulic}. Since we assume for simplicity the
former case, then $\Sigma _{C}$ is assumed in the form (after the
$\xi $-integration)
\begin{equation}
\Sigma _{C,n,\varphi }=\pi T\sum_{m}\frac{\lambda _{C,\varphi
}(\omega
_{n}-\omega _{m})\tilde{\omega}_{m,\varphi }}{\sqrt{\tilde{\omega}%
_{m,\varphi }^{2}+\tilde{\Delta}_{m,\varphi }^{2}}},  \eqnum{4}
\end{equation}
where
\begin{equation}
\lambda _{C,\varphi }(n-m)=2\int_{0}^{\infty }d\Omega
\frac{S_{C,\varphi }(\Omega )\Omega }{\Omega ^{2}+(\omega
_{n}-\omega _{m})^{2}}.  \eqnum{5}
\end{equation}
ARPES predicts that $\mathop{\rm Im}%
\Sigma _{C,\varphi }(\omega )\sim -\pi \lambda
_{C,\varphi}\omega/2$ at $ T<\omega <\omega _{C}$ which we
reproduce by taking $S_{C,\varphi }(\omega )=A_{C,\varphi }\Theta
(\mid \omega \mid -T)\Theta (\Omega _{C}-\mid \omega \mid )$.
$A_{C,\varphi }$ is normalized in such a way to obtain $\lambda
_{C,\varphi }<0.4$.

$Eq.2$ contains the Hartree-Fock pseudopotential $\mu ^{\ast
}(\varphi ,\varphi ^{\prime \prime })$ \cite{allen} which is a
dull function of ($\varphi ,\varphi ^{\prime }$), i.e. $\mu ^{\ast
}(\varphi ,\varphi ^{\prime
})=\mu _{0}^{\ast }>0$. It maximizes $T_{c}$ when $\tilde{\Delta}%
_{n,\varphi }$ is d-wave like. $\tilde{\Delta}%
_{C,n,\varphi }$ in $Eq.(2)$ describes other effects of the
Coulomb interaction to pairing which are unknown. The theory
\cite{allen} gives $\tilde{\Delta}_{C}=(\Gamma _{e}/\varepsilon
_{e})V_{C}G\Gamma _{e}$ (with $Z(k,\omega _{n})>1$ and $(\Gamma
_{e}/\varepsilon _{e})>0$), then $\tilde{\Delta}_{C}$ is modeled
according to ARPES. For instance, the SF approach assumes a
phenomenological form for $\tilde{\Delta}_{C}({\bf k},\omega
_{n})$ and for $\Sigma _{C}$,
which depend on the dynamical spin susceptibility $%
\chi _{spin}$. Since $\chi _{spin}({\bf q},\omega)$ is peaked at
${\bf Q}=(\pi ,\pi )$ this term is repulsive and favors d-wave
pairing. However, the SF proponents assume an unjustifiable large
coupling $g_{sf}\approx (0.5-0.65)$ $eV$ (i.e. $\lambda
_{sf}\approx 2-3$). The analyzes of various experiments gives
$g_{sf}\lesssim 0.1$ $eV$ (i.e. $\lambda _{sf}<0.2)$ \cite
{kulic}, which is confirmed by
ARPES \cite{lanzara}, \cite{damascelli}, \cite{shen2} for $\mathop{\rm Im}%
\Sigma _{C}({\bf k},\omega )$ at $\omega >\omega _{ph}^{\max }$.
It gives also small $\lambda _{sf}(<\lambda _{C}<0.4$). Although
the SF term in $\tilde{\Delta}_{C,n,\varphi }$
is much smaller than the EPI it is important (together with $%
\mu _{0}^{\ast }$ and other Coulomb terms) in triggering SC from
s- wave to d-wave pairing \cite{kulic2}, \cite{kulic}. In
$Eqs.(1-2)$ non-magnetic impurities are included and strong
correlations \cite{kulic} induce the FSP in the impurity
scattering matrix, being $t(\varphi ,\varphi \prime ,\omega )\sim
\gamma _{c}^{2}(\varphi {\bf -}\varphi ^{\prime })$. In the
leading order
one has $t(\varphi ,\varphi \prime ,\omega )\sim \delta (\varphi {\bf -}%
\varphi ^{\prime })$, thus not affecting (any kind of) pairing. In
reality they are pair-breaking for d-wave pairing and the next to
leading term is necessary. It is characterized by two scattering
rates, $\gamma _{1,\varphi }$ and $\gamma _{2,\varphi }$, where
$\gamma _{1,\varphi }-\gamma _{2,\varphi }>0$. $\gamma _{1,\varphi
}=\gamma _{2,\varphi }$ mimics the extreme forward scattering and
$\gamma _{2,\varphi }=0$ means the isotropic pair-breaking
scattering.

{\it Quasiparticle renormalization in nodal and anti-nodal
directions - }The quasiparticle energy $\omega (\xi_{\bf k})$ is
the pole of the retarded Green's function. For numerical
calculations we assume a Lorenzian shape for $\alpha _{ph,\varphi
}^{2}F_{\varphi }(\Omega )=A_{ph,\varphi }/(W^{2}+(\Omega -\omega
_{ph,\varphi })^{2}$. The parameters $A_{ph,\varphi }$, $W$
are chosen so that $\lambda _{ph,\varphi }(0)=\lambda _{ph,\varphi }$, where $%
\lambda _{ph,\varphi }$ is an effective EPI coupling at the point $%
\varphi $. Since our aim is a qualitative explanation of the {\em
ARPES shift-puzzle} we perform calculations only for moderate
couplings $\lambda _{ph}=1$, $\lambda _{C}=0.3$ in both, the nodal
and
anti-nodal direction. In fact their real values might be larger, i.e. $%
\lambda _{ph}\lesssim 2$, $\lambda _{C}\approx 0.4$. $Eqs.(1-5)$
are local (angle-decoupled) on the Fermi surface, i.e. $\Delta
_{n,\varphi }$ is locally ''s-wave SC'' and globally d-wave
pairing, which is also manifested for\ more realistic
interactions, whenever $\delta \varphi _{w}\ll \pi $ - see
\cite{kulic}.

\textbf {(1)} $\omega _{kink}^{(70)}$-{\em kink in the nodal direction}$%
(\varphi =\pi /4$) - The kink at $\omega _{kink}^{(70)}\approx 70$
$meV$ in $\omega (\xi_{\bf k})$ means that the quasiparticles
moving along the nodal direction interact with various phonons
with frequencies up to 70 meV  \cite{carbotte}, i.e. $\alpha
_{ph,\pi /4}^{2}F_{\pi /4}(\Omega )\neq 0$ for $0<\Omega \lesssim
70$ $meV$. Since it is unknown, a Lorenzian shape centered at
$\omega _{ph}^=70$ $meV$ is assumed. In this case the theory
predicts more singularity-like \cite{carbotte} than the observed
kink-like behavior. $\Delta _{\pi /4}(\omega )=0$ and $Eq.(1)$
imply that $\omega (\xi_{\bf k})$ is equal in the N and in the SC
state, as it is shown in $Fig.1a$. It confirms that {\em the kink
in the nodal direction is not shifted in the SC state} - in a
qualitative agreement with ARPES findings \cite{lanzara}.

\begin{figure}[tbp]
\includegraphics*[width=8cm]{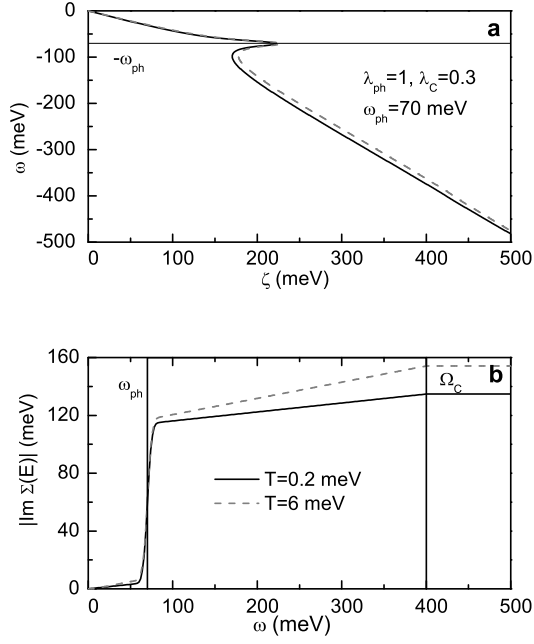}
\caption{{\bf a} - The quasiparticle-spectrum $\omega (\xi_{\bf
k})$ and {\bf b} - the imaginary self-energy $\mathop{\rm Im}
\Sigma (\xi =0,\omega)$
in the nodal direction ($%
\varphi =\pi /4$) in the SC ($T=0.2$ $meV$) and N ($T=6$ $meV$)
state. $\Omega _{C}=400$ $meV$ is the cutoff in $S_{C}$.}
\end{figure}
The realistic phonon spectrum will smear the theoretical
singularity in $\omega (\xi_{\bf
k})$ - seen in $Fig1a$. In $Fig.1b$ is shown $\mathop{\rm Im}%
\Sigma (\xi =0,\omega )$, where a qualitative similarity with
ARPES \cite{lanzara} is obvious. For $\omega _{ph}^{(70)}<\omega
<\Omega _{C}$ the linear term
$\mid \mathop{\rm Im}%
\Sigma _{C}(\xi =0,\omega )\mid \sim \omega $ is discernable,
while near $\omega _{ph}^{(70)}$ $\mathop{\rm Im}%
\Sigma (\xi =0,\omega )$ is steeper due to $\lambda _{ph}(=1)\gg
\lambda _{C}(=0.3)$.

\textbf {(2)} $\omega _{sing}^{(40)}${\em -singularity in the
anti-nodal direction} ($\varphi \approx \pi /2$)- The singularity
(not the kink) at $\omega _{sing}^{(40)}$ in $\omega (\xi_{\bf
k})$ in the anti-nodal
direction is observed in ARPES in the N and SC state of $%
La_{2-x}Sr_{x}CuO_{4}$ and $BISCO$ \cite{shen2}, which means that
the quasiparticle moving in the anti-nodal direction interact with
a narrower phonon spectrum centered around $\omega
_{ph}^{(40)}\approx 40$ $meV$. So, the assumed (by us) the
Lorenzian shape for $\alpha _{ph,\varphi \approx \pi
/2}^{2}F_{\varphi \approx \pi /2}(\Omega )$, centered at $\omega
_{ph} \approx 40$ $meV$, is acceptable approximation. Since
$\Delta _{\pi /2}(\omega )=\pm \Delta _{0}$ then $Eq.(1)$ gives
that $\omega (\xi_{\bf k})$ in the N-state is singular at $\omega
_{sing}=\pm \omega _{ph}^{(40)}$, while in the $SC$ state {\em it
is shifted} to $\omega _{sing}^{(40)}=\pm (\omega
_{ph}^{(40)}+\Delta _{0})$. This is confirmed by numerical
calculations shown in $Fig.2a$ - $\omega
(\xi_{\bf k})$, and in $Fig.2b$ - $\mathop{\rm Im}%
\Sigma (\varphi ,\omega )$, for $\lambda _{ph}=1$ and $\lambda
_{C}=0.3$.

\begin{figure}[tbp]
\includegraphics*[width=8cm]{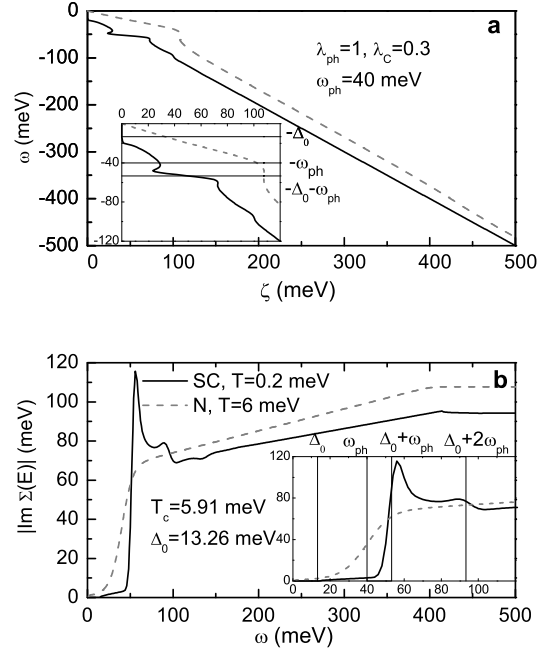}
\caption{{\bf a} - The quasiparticle-spectrum $\omega (\xi_{\bf
k})$ and {\bf b} - the imaginary self-energy $\mathop{\rm Im}
\Sigma (\xi =0,\omega )$ in the anti-nodal direction ($\varphi
=0;\pi /2$) at $T=0.2$ $meV$ in in the SC ($T=0.2$ $meV$) and N
($T=6$ $meV$) state.}
\end{figure}

The $\omega _{sing}^{(40)}$ singularity is shifted in the SC
state, contrary to the nodal kink at $\omega _{kink}^{(70)}$ which
is not. So, the different shifts of $\omega _{kink}^{(70)}$ and
$\omega _{sing}^{(40)}$ in the SC state is a {\em direct
consequence of the forward scattering peak in the charge
scattering processes}. Since we assume a rather narrow phonon
spectrum (centered around $\Omega _{ph}$) the behavior of
$\mathop{\rm Im} \Sigma (\xi =0,\omega )$ at $\omega \ll \Omega
_{ph}$ is due to the Coulomb interaction - the small tails in
$Fig1b$ and $Fig2b$.

\textbf {(3)} {\em ARPES dip-hump structure } - The FSP-model
explains qualitatively the dip-hump structure in $A(\varphi
,\omega )(=-\frac{1}{\pi }
\mathop{\rm Im}%
G(\varphi ,\omega ))$. The latter was observed recently in ARPES
\cite{damascelli}, where the dip is very pronounced in the SC
state. In $Fig.3a$ it is seen that the dip-hump structure is
realized (also in the presence of impurities) already for a
moderate coupling $\lambda _{ph}=1$ in the N state, while it is
more pronounced in the SC state.
\begin{figure}[tbp]
\includegraphics*[width=8cm]{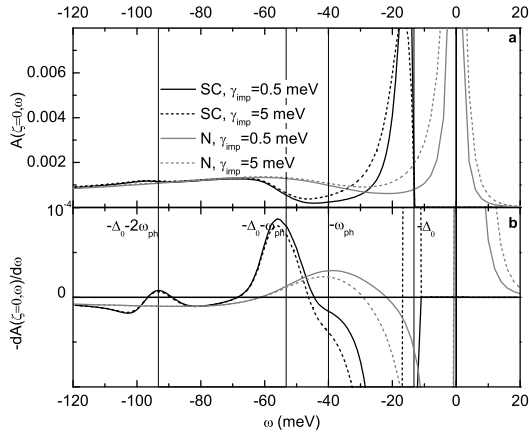}
\caption{{\bf a} - The spectral function $A(\xi =0,\omega )$ and
{\bf b} - $-dA(\xi =0,\omega )/d\omega $ in the anti-nodal
direction in the SC ($T=0.2$  $meV$) and N ($T=6$ $meV$) state for
various non-magnetic impurity scattering rate $\gamma_{1}$ and
$\gamma_{2}=0$; $\lambda _{ph}=1$, $\lambda _{C}=0.3$.}
\end{figure}
$A(\omega )$ is appreciable narrowed in the SC state. It seems
that the dip-energy can not be attached to the (shifted) phonon
energy at $\omega_{ph} =40$ $meV$ only, since the maxima of
$-dA/d\omega $ have more universal meaning (than the minima of
$A$) - see $Fig.3b$. The maxima in -$dA/d\omega $ appear near the
energies ($-\Delta _{0}-n\omega_{ph} $). The calculations give the
dip structure also in the anti-nodal density of states $N_{\pi
/2}(\omega )$ (not shown) already for $\lambda _{ph}=1$, which is
much more pronounced for larger $\lambda _{ph}$.

{\it Discussion and conclusions} - In obtaining $Eqs.(1-5)$ in the
FSP model several approximations are made: ({\bf i}) the charge
scattering spectral functions are assumed to be $\sim \delta
(\varphi -\varphi ^{\prime })$. This extreme limit is never
realized in nature, but for the self-energy it
is a good starting point. The finite $\delta \varphi _{w}$ effects (but $%
\delta \varphi _{w}\ll \pi $) will not change the qualitative
picture but only the quantitative one \cite{kulic}. In previous
studies \cite{kulic} the EPI spectral function was treated in the
extreme momentum FSP limit, were they were proportional to $\delta
({\bf k}-{\bf k}^{\prime })$ - the MFSP model. The latter resolves
the {\em ARPES shift-puzzle} too, but its self-energy is more
singular than in the FSP model. In reality the
spectral functions are broadened in the interval $\mid {\bf k}-{\bf k}%
^{\prime }$ $\mid <\delta k_{c}\ll k_{F}$ and the effects of
finite level-spacing ($\sim 1/N$) in ${\bf k}$-space are absent;
({\bf ii}) the Migdal theory is assumed to hold and vertex
corrections due to the EPI are neglected. However, in the
FSP-model vertex corrections may be important for $\lambda
_{ph}<1$ \cite{pietro}, by increasing $T_{c}$ significantly; ({\bf
iii}) the role of the Coulomb repulsion in the anomalous
self-energy $Eq.(2)$ is unknown, but since the calculation of
$T_{c}$ was not the (main) purpose of this paper and because ARPES
and other experiments suggest that $\lambda _{C}\ll \lambda _{ph}$
we have omitted its contribution to the gap equation.

In conclusion, we analyze the quasiparticle self-energy effects
for HTSC oxides in the theoretical model with the pronounced
forward scattering peak in the electron-phonon interaction (which
dominates in HTSC), Coulomb interaction and impurity scattering -
the FSP model. The different shifts of the nodal kink (at 70 meV)
and anti-nodal singularity (at 40 meV) in the superconducting
state, which are observed in the ARPES experiments on HTSC oxides
\cite{lanzara}, \cite{shen2}, are explained by the FSP model in a
consistent and unique way. However, a quantitative refinement of
the FSP\ model is needed, which must take into account realistic
phonon and band structure, bi(multi)layer structure, less (than
delta-function) singular spectral functions, etc.

{\bf Acknowledgment -} M. L. K. thanks Prof. Ulrich Eckern and
Drs. Igor and Lila Kuli\'{c} for support.

\end{document}